\documentclass{elsart}

\usepackage{graphicx}

\usepackage{amssymb}

\begin{document}

\begin{frontmatter}

\title{Sliding susceptibility of a rough cylinder on a rough inclined perturbed surface\thanksref{support}}
\thanks[support]{Work supported in part by CNPq, FINEP and PRONEX (Brazilian government agencies).}

\author{V. P. Brito}\and{\author{R. F. Costa}} 
\address{
Departamento de F\'{\i}sica, Universidade Federal do Piau\'{\i}, 64049-550, Teresina, PI, Brazil.
}
\author{M. A. F. Gomes\corauthref{mafg}}
\corauth[mafg]{Corresponding author. 
fax: +55-81-3271-0359}
\ead{mafg@ufpe.br}
\address{
Departamento de F\'{\i}sica, Universidade Federal de Pernambuco,
50670-901, Recife, Brazil.
}
\author{E. J. R. Parteli}
\address{
Institut fuer Computeranwendungen, ICA-1, Pfaffenwaldring 27, D-70569, Stuttgart, Germany.
}

\begin{abstract}
A susceptibility function ${\chi}(L)$ is introduced to quantify some aspects of the intermittent stick-slip dynamics of a rough metallic cylinder of length $L$ on a rough metallic incline submitted to small controlled perturbations and maintained below the angle of repose. This problem is studied from the experimental point of view and the observed power-law behavior of ${\chi}(L)$ is justified through the use of a general class of scaling hypotheses. 

\end{abstract}

\end{frontmatter}

\section{Introduction}
Several dynamic processes of physical \cite{liu}, geologic \cite{scholz}, and technological \cite{bowden} interest involve superposed interacting rough surfaces able to display stick-slip motion \cite{heslot}. It has been discovered in the last few years that the intermittent sliding or stick-slip dynamics of a rough solid cylinder on a rough inclined surface submitted to small controlled perturbations is a fluctuation phenomenon characterized by non-trivial spatiotemporal scaling laws \cite{brito,gomes}, and complex critical exponents \cite{parteli} if the inclination is below the angle of repose. In the usual stick-slip dynamics a solid is pulled at a constant driving velocity; in the experiments discussed in this paper, on the other hand, the stick-slip dynamics appears as a consequence of a different mechanism: to start the slip we resort to small mechanical perturbations on an inclined surface below the angle of repose. A solid body on a perturbed incline is an example of a nonequilibrium system receiving an incoming energy flow. If energy is continuously injected into nonequilibrium systems, a complex sequential response characterized by time series of events of all sizes is often observed. Besides sliding blocks on inclines, other important examples of systems and phenomena associated with a similar type of temporal response include piles of sand and other granular materials \cite{held,miller}; acoustic emission from volcanic rocks and microfracturing processes in general \cite{diodati,petri}; interface depinning in magnetic systems \cite{urbach}; stick-slip motion in lubricated systems \cite{demirel}, and turbulence \cite{frisch}, among others. 

In this paper, the sliding susceptibility ${\chi}(L)$ of a rough metallic cylinder of length $L$ placed on a rough inclined surface submitted to small controlled perturbations is introduced and examined. To obtain this response function, we performed a class of very time consuming experiments in which we count the number of perturbations on the incline that give origin to a certain number of sliding events of the cylinder. The total number of perturbations in our experiments exceeded $56,000$, corresponding to $6,000$ sliding events. Before and after each perturbation, the position of the cylinder along the incline as well as other variables of interest must be controlled. We find that ${\chi}(L)$ presents scaling symmetry \cite{feder} in a number of physical situations and the origin of this behavior is investigated.

The outline of the article is as follows. The experimental details are described in Sec. 2; and in Sec. 3 we report the experimental data, and discuss our principal results. Also in Sec. 3, we present scaling hypotheses to explain our findings. A summary of our major conclusions and future prospects are found in Sec. 4.

\section{Experimental details}

\begin{figure}
\begin{center}
\includegraphics*[width=0.75\columnwidth]{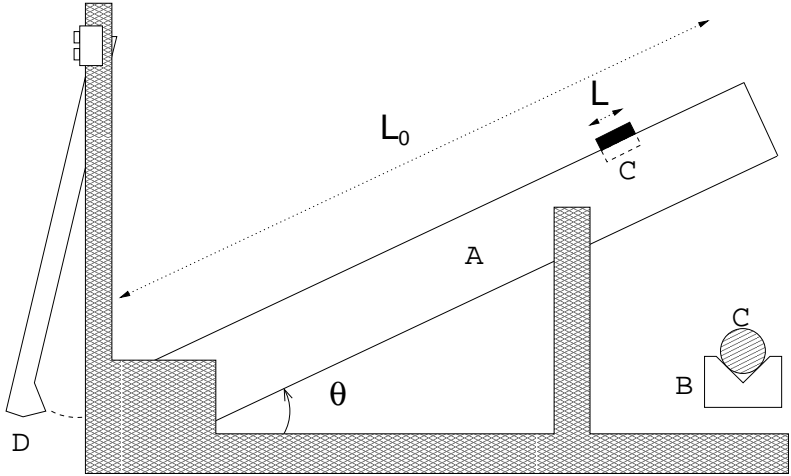}
\caption{Schematic diagram of the experimental apparatus: the incline (A) has a $L_0=1,500$mm long groove (the cross section is shown in B) and is rigidly maintained at an inclination $\theta$ with the horizontal. The cylinders (C) have lengths varying from $L=5$mm to $L=1000$mm. The hammer of $m=175$g or $75$g (D) impacts the base of the incline with a velocity close to $50\,{\mbox{cm}}\,{\mbox{s}}^{-1}$. See text for detail.}
\label{fig:experiment}
\end{center}
\end{figure}

The basic apparatus used to obtain the experimental data is outlined in Fig. {\ref{fig:experiment}}: it consists of a rigid V-shaped anodized aluminum chute made of corner plate of $5\,$mm of thickness and with $90^{\circ}$ angular aperture symmetrically disposed with respect to the vertical plane. The chute was rigidly maintained with an inclination $\theta$ with respect to the horizontal, and it is supplied with an articulated hammer of mass $m=175$g or $75$g which hits the base of the chute with a controlled (fixed) velocity close to $50$cm/s at the moment of the impact. The system is mounted on a 200kg table isolated from mechanical vibrations. On the chute (with an effective length $L_0=1,500$mm) is placed a metallic cylinder of length $L$ (5mm to $1,000$mm), and the system chute $+$ cylinder operates below the critical angle of repose ${\theta}_{\mathrm{c}}$=${\tan}^{-1}{\mu}_{\mathrm{s}}$, where ${\mu}_{\mathrm{s}}$ is the coefficient of static friction of the cylinder on the chute. Two types of chute, hammer, and metallic cylinders were used as will be specified below. In the regime of inclination used in the experiments, the series of induced sliding events are intermittent, i.e. a fluctuating number of many controlled perturbations of the hammer is necessary to induce a single sliding event of the cylinder. Each constant perturbation on the chute is associated with a time unit. The number of impacts of the hammer to obtain each sliding was recorded and from this data we performed the statistical analysis discussed in this paper. In all experiments the inclination was in the interval $12^{\circ}$ to $18^{\circ}$, with $16^{\circ} \leq {\theta}_{\mathrm{c}} \leq 32^{\circ}$, and the reduced angle $({\theta}_{\mathrm{c}} - \theta)/{\theta}_{\mathrm{c}}$ was in the interval $0.19$ to $0.37$ (see below for details). The sliding experiments were {\it{polarized}} in the sense that the cylinders were always placed on the chute in a same axial orientation. Two types of experiments were made with the azimuthal orientation of the cylinder fixed or not. In the first case, rotations of the cylinder around its symmetry axis are blocked, and as a consequence the interacting surfaces in contact were the same during all the experiment. The sliding spatial resolution is $0.5$mm, and the angular resolution for the angle of inclination is $0.1^{\circ}$. Only sliding events with a length equal or larger than a threshold of $1$mm are recorded for statistical purposes. The average sliding length in the experiments varies typically in the interval of $5$ to $10$mm. The experiments were realized in a small sealed room at a temperature of $24\,(\pm 1)^{\circ}$C, and $59\,(\pm3)\%$ of humidity. Chute and cylinder operated free of lubrication and both surfaces were submitted to a periodic cleaning with ethanol and cotton wool. The experimental data obtained can be classified in four different groups, as follows. 

{\it{Group G1}} consists of 9 time series of sliding events obtained in 1996 for massive cylinders with fixed diameter $\phi = 9.6$mm, and 9 different lengths $L$(mm) = 5, 10, 20, 50, 100, 200, 300, 500, and 1000. Here and in all the other groups mentioned below, a series of sliding events is planned to generate a total of 100 sliding events. Both chute and cylinder have nominal deviations of the profile from a smooth surface in the range between about $0.1 {\mu}$m and $1 {\mu}$m \cite{thewlis}. The mass of the hammer used to introduce the perturbations on the rigid V-shaped chute was $m_1=175$g. The total duration of these time series varies from 364 (for $L=1000$mm) to $6,979$ (for $L=5$mm) time units or hammer impacts. In this group the azimuthal orientation of the cylinder is held fixed. The {\it{average}} (on the nine values of $L$ previously mentioned) of the reduced angle of inclination in this group is $\left<{\theta}_{\mathrm{r}}\right> = \left<{({\theta}_{\mathrm{c}}-{\theta})/{\theta}_{\mathrm{c}}}\right> = 0.26 \pm 0.07$.

{\it{Group G2}} consists of 27 time series obtained with the same set of cylinders and chute of G1 (here there are 3 independent time series associated with each one of the 9 values of $L$), but the data were collected in 2002, i.e. this group incorporates complex aging effects as for example the oxidation of the several interacting surfaces. Two relevant aspects in this group are: (i) before the beginning of the experiments, the aluminum cylinders and the chute had to be cleaned firstly with a fine steel wool, and after with water and neutral soap. (ii) The mass of the hammer used in the experiments in G2 was $m_2 = 75$g. It is expected on the basis of these two modifications that the signal to noise ratio is somewhat inferior for G2 as compared with G1. The effect of these modifications will be discussed in Sec. 3. Here, differently from G1, both chute and cylinder have nominal deviations of the profile from a smooth surface in the range between about $1{\mu}$m and $10{\mu}$m \cite{thewlis} after the use of the steel wool. In this group, the experiments maintained the same interface cylinder-chute by preventing azimuthal rotations of the cylinder around its longitudinal axis. The average of the reduced angle of inclination in this group is the same as in G1: $\left<{{\theta}_{\mathrm{r}}}\right> = 0.26 \pm 0.07$.

{\it{Group G3}} refers to 17 time series of sliding events for the same cylinders and chute of G2, but the length $L$ of the cylinders were restricted to the values $L$(mm) $=$ 100, 200, 300, 500, and 1000. In the experiments grouped in G3, the control to maintain the cylinders free of rotations around their longitudinal axes was relaxed, i.e. in this case the cylinder-chute interface can vary continuously. Especially the cylinders with the smaller lengths are more sensitive to execute small rotations around their longitudinal axes. The hammer used in G3 was the same used in G2. In this group ${\theta}_{\mathrm{r}}$ (not an average) is fixed at $0.37$.

{\it{Group G4}} consists of 9 time series of sliding events obtained with a new set of aluminum cylinders and rigid aluminum chute (also 1500mm long), with the same type of surface finishing used to obtain the data in G1. In this group we have used hollow cylinders of 10mm external diameter and with 1mm wall width. The cylinders are free to rotate around their symmetry axes as in G3, and the hammer is the same used to obtain the data in G2 and G3. The length of the cylinders used in this case and the associated number of equivalent time series were, respectively 1000mm/three series, 500mm/two series, 200mm/two series, and 100mm/two series. In this group ${\theta}_{\mathrm{r}}$ is fixed at $0.25$ for all the values of $L$. 

Before to conclude this Section, we would like to justify briefly the choice of some parameters in our experiments. Six years after the acquisition of the data grouped in G1, we decided to test the robustness of this type of sliding experiment. As a consequence, the experiment with G2 was planned to include more time series: performing three experiments for each value of the length $L$ was considered a compromise between reliability and limitation of time. After a long period to obtain the data grouped in G2 we were forced to restrict the variability of the length of the cylinders for G3 and G4 (from 9 values of $L$ in G1 and G2 to 5 values of $L$ in G3, and 4 values in G4) while maintaining the approximate number of time series (2-3) for each value of length. This decision was taken to avoid a pronounced abrasion of the chute. To compensate this limitation, we increased our control on the reduced angle of inclination.

\section{Results and discussion}

We have measured in the experiments the number of impacts of the hammer, $T_N(L)$, necessary to obtain $N$ sliding events with size $\lambda \geq 1$mm, for a cylinder with length $L$. To quantify the propensity or susceptibility of the cylinder to slide on the chute we introduce the sliding susceptibility ${\chi}_N(L)$ defined as the ratio of the number of sliding events or failures to stick, $N$, to the corresponding number of trials to move the cylinder through the controlled perturbations on the chute, $T_N(L)$, that is
\begin{equation}
{\chi}_N(L) = N/T_N(L).
\end{equation}
Thus, Eq. (1) is an {\it{average}} quantity given by the reciprocal of the number of perturbations necessary to obtain one sliding event. If a large number of perturbations is necessary to obtain only a small number of sliding events, the corresponding susceptibility will be small; on the contrary, if each perturbation of the hammer is followed by a sliding event, the susceptibility reaches its maximum value ${\chi}_{\mathrm{max}}=1$.

\begin{figure}
\begin{center}
\includegraphics*[width=0.9\columnwidth]{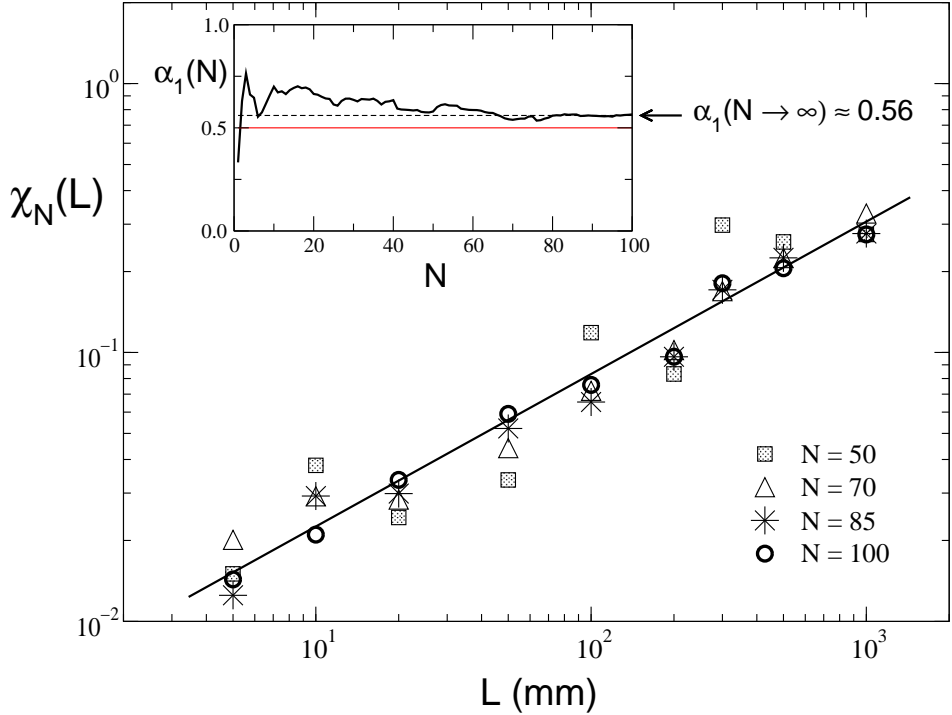}
\caption{Log-log plot of the experimental sliding susceptibility ${\chi}_N(L)$, for $N=50$, 70, 85, and 100, and $5 \leq L$(mm)$\,\leq 1000$, for group G1 defined in Section 2. The straight line in the main figure represent the power-law best fit associated with these particular values of $N$. Our overall estimate from these data is the scaling relation $\chi \sim L^{0.55 \pm 0.05}$. The inset shows the exponent ${\alpha}_1(N)$ obtained from the adjust ${\chi}_N \sim L^{{\alpha}_1(N)}$, for $N$ varying from 1 to 100. The asymptotic value of the scaling exponent seems to be ${\alpha}_1(\infty) = 0.56$. See text, Sec. 3 for detail.}
\label{fig:T_N(L)_G1}
\end{center}
\end{figure}

Figure {\ref{fig:T_N(L)_G1}} shows ${\chi}_N(L)$ for $N=$ 50, 70, 85, and 100, for nine values of $L$ in the interval $5 \leq L$(mm)$\,\leq 1000$, for the data coming from the group G1 described in Section 2. ${\chi}_N(L)$ in this case increases as the power-law $L^{{\alpha}_1(N)}$, along $2.3$ decades of variability in $L$. The exponent ${\alpha}_1(N)$ as a function of $N$, obtained after a best fit analysis of the experimental data for $N$ varying from 1 to 100 is given in the inset of Fig. {\ref{fig:T_N(L)_G1}}. Its asymptotic value seems to be ${\alpha}_1(\infty)=0.56$. From Fig. {\ref{fig:T_N(L)_G1}} we can conclude that the scaling relation for ${\chi}_N(L)$ is reasonably robust and independent of $N$ within typical statistical fluctuations of $5\%$ to $10\%$ in the exponent $\alpha$. Our overall estimate for this exponent from the data in Fig. {\ref{fig:T_N(L)_G1}} is $\alpha = 0.55 \pm 0.05$. For $N=100$, the average number of perturbations necessary for a single sliding event of the 5mm cylinder is close to 70 (the experimental point associated with $L=5$mm in Fig. {\ref{fig:T_N(L)_G1}} corresponds to an average on just $6,979$ perturbations), whereas for the longest cylinder of 1000mm, the corresponding average number of perturbations is close to $3.6$ (the experimental point associated with $L=1000$mm in Fig. {\ref{fig:T_N(L)_G1}} corresponds to an average on just 364 perturbations). From now on we will focus on ${\chi}_N(L)$ for $N=100$, because for this particular value of $N$ the susceptibility function appears to be sufficiently close to its thermodynamic limit. Indeed, a detailed analysis (not shown in Fig. {\ref{fig:T_N(L)_G1}}) indicates that the values of ${\chi}_N(L)$ for $L$ fixed seem to approach a limit as $N$ increases from 50 to 100. Moreover, for $N=100$ the exponent $\alpha$ also seems to have already reached its thermodynamic-limit value, as we can conclude from the inset of Fig. {\ref{fig:T_N(L)_G1}}. For brevity we will then use the notation ${\chi}_{N=100}(L) \equiv {\chi}(L)$. 

\begin{figure}
\begin{center}
\includegraphics*[width=.8\columnwidth]{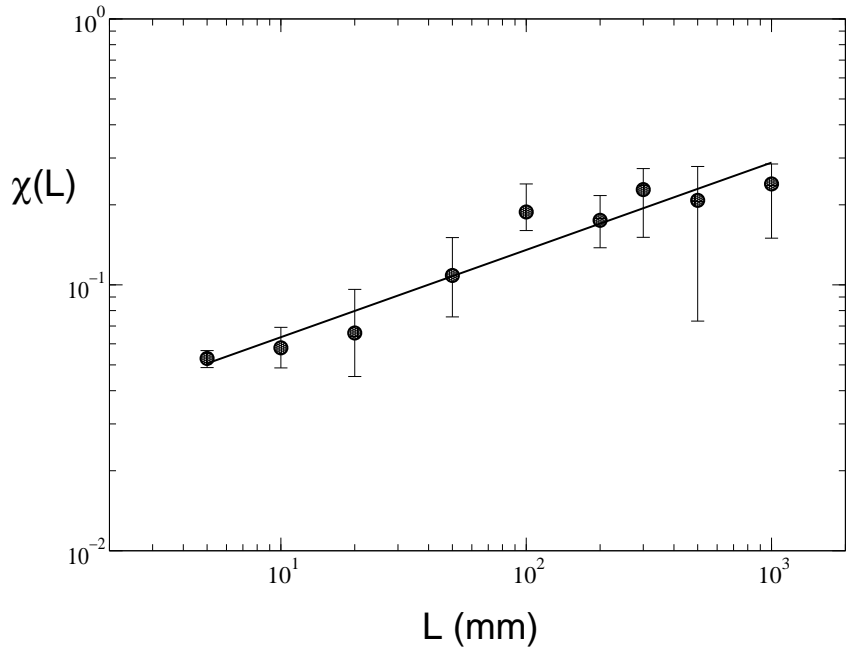}
\caption{The same as in Fig. {\ref{fig:T_N(L)_G1}} but for group G2: the power-law fit to the experimental data gives $\chi(L) \sim L^{0.33 \pm 0.04}$. Each point in this figure represents an average on three independent time series. See text, Secs. 2 and 3 for detail.}
\label{fig:T_N(L)_G2}
\end{center}
\end{figure}

In Fig. {\ref{fig:T_N(L)_G2}} it is shown ${\chi}(L)$ for G2 together with a power-law fit (continuous line) to the experimental data; here $\chi(L) \sim L^{{\alpha}_2}$, with ${\alpha}_2 = 0.33 \pm 0.04$. Each point in Fig. {\ref{fig:T_N(L)_G2}} for a fixed value of $L$ represents an average on three independent time series. In this case, the reduction of near $60\%$ in the mass of the hammer as compared with G1 is accompanied by a reduction close to $40\%$ in the scaling exponent. This reduction in the impact energy, as well as the alterations in the surfaces introduced by the cleanness of the cylinders and chute with steel wool and the aging effects introduced along the interval of six years between the experiments grouped in G1 and G2 (See Sec. 2) lead to a reduction in the signal/noise relation which is apparently materialized in the reduction of the exponent $\alpha$ and in the fluctuations in the power-law scaling (in particular for $L=100$mm). 

\begin{figure}
\begin{center}
\includegraphics*[width=.8\columnwidth]{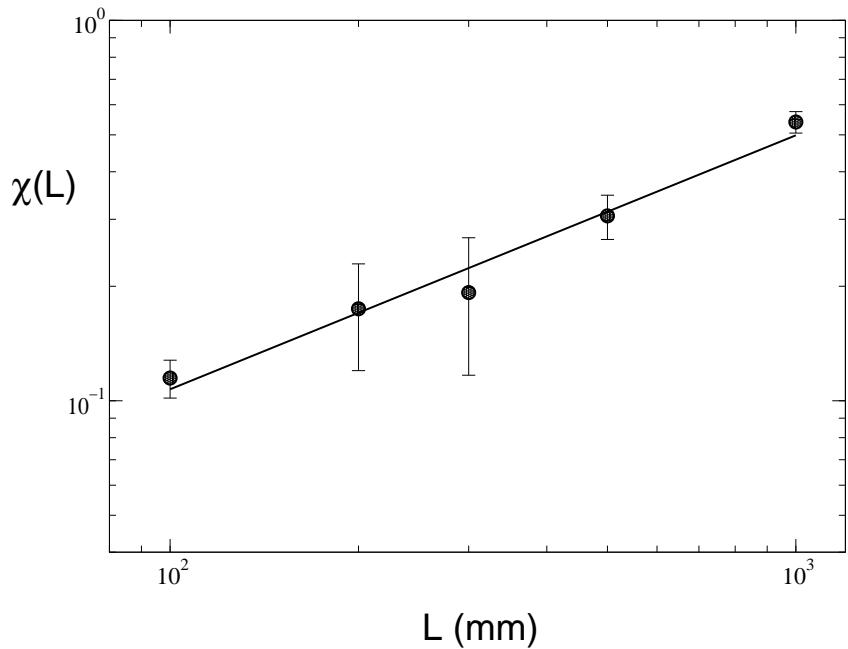}
\caption{Plot of the sliding susceptibility for G3; the best fit to the data gives $\chi \sim L^{0.67 \pm 0.06}$, for $100 \leq L$(mm)$\, \leq 1000$. As in Fig. {\ref{fig:T_N(L)_G2}}, each point represents an average on independent time series. See text, Secs. 2 and 3 for detail.}
\label{fig:T_N(L)_G3}
\end{center}
\end{figure}

The experimental dependence of $\chi(L)$ for G3 is shown in Fig. {\ref{fig:T_N(L)_G3}}; this susceptibility increases as $L^{{\alpha}_3}$, with ${\alpha}_3 = 0.67 \pm 0.06$ for $100 \leq L$(mm)$\leq 1000$. Each point in Fig. {\ref{fig:T_N(L)_G3}} for a fixed value of $L$ represents an average on two to four independent time series. This average is important to obtain a good scaling behavior through the elimination of fluctuations associated with the tendency of the shortest cylinders to rotate around their symmetry axes. 

\begin{figure}
\begin{center}
\includegraphics*[width=.8\columnwidth]{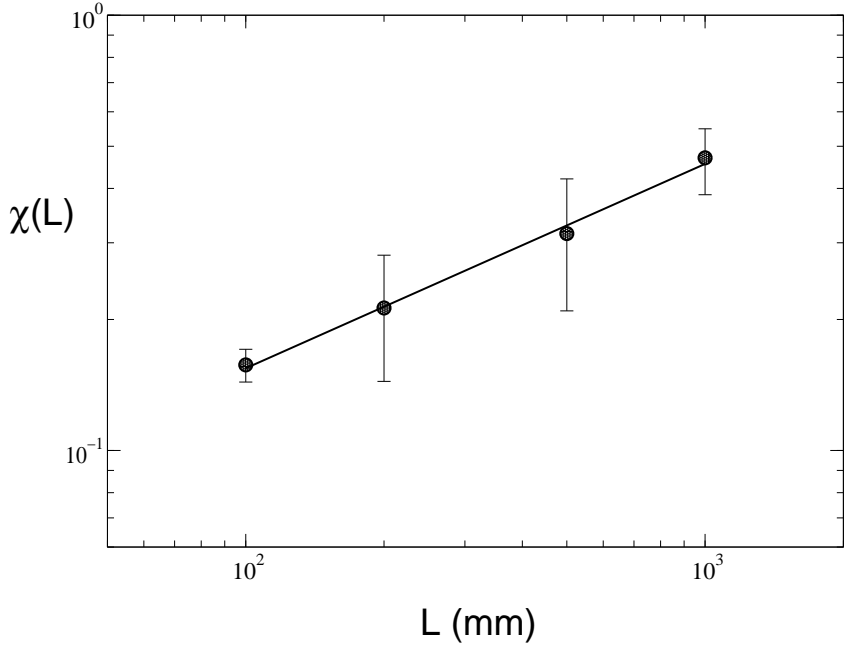}
\caption{Plot of the sliding susceptibility for hollow cylinders free to rotate around the longitudinal axis (group G4). In this case $\chi(L) \sim L^{0.48 \pm 0.02}$. As in Figs. {\ref{fig:T_N(L)_G2}} and {\ref{fig:T_N(L)_G3}}, each point in this plot represents an average on independent time series. See text, Secs. 2 and 3 for detail.}
\label{fig:T_N(L)_G4}
\end{center}
\end{figure}

For hollow cylinders (G4) free to rotate around the longitudinal axis as with G3 we have obtained the data presented in Fig. {\ref{fig:T_N(L)_G4}}, with $\chi(L) \sim L^{{\alpha}_4}$, ${\alpha}_4 = 0.48 \pm 0.02$. In this case, to find a power-law fit similar to those appearing in Figs. {\ref{fig:T_N(L)_G1}} to {\ref{fig:T_N(L)_G3}}, we need again to perform an average on similar time series to attenuate the fluctuations due to rotations. As noticed in Sec. 2, a total of 9 time series were used to get Fig. {\ref{fig:T_N(L)_G4}}: average on 3 series for cylinders with $L=1000$mm; average on 2 time series for $L=500$mm; average on 2 series for $L=200$mm, and average on 2 series, for $L=100$mm. 

The first physical aspect that we consider to explain the experimental results reported in this work is the effect of the vibrations induced on the chute and on the cylinder as a consequence of the impacts of the hammer. A simple heuristic argument predicts that the susceptibility Eq. (1) which is proportional to the probability of the cylinder to slip after a perturbation would scale linearly with the total number $\nu$ of longitudinal/transversal/torsional vibration modes of the cylinder:
\begin{equation}
\chi \sim \nu.
\end{equation}
This is a reasonable assumption because (i) cylinders with more vibration modes tend to disengage more easily from the contacts that tend to maintain them fixed with respect to the chute, and (ii) the role of the chute on Eq. (2) can be eliminated because it is the same for cylinders of different lengths. On the other hand, for a thin rod-like system uniform along its length $L$ we expect that
\begin{equation}
\nu \sim {(L/{\lambda}_{\mathrm{min}})}^{{\alpha}_0} \sim {(L/{\phi})}^{{\alpha}_0}, \ \ \alpha \rightarrow {\alpha}_0 = 1,
\end{equation}
where ${\lambda}_{\mathrm{min}} \cong \phi$ ($=$ diameter of the rod/cylinder) is the minimum (cutoff) wavelength allowed \cite{thomson}. If Eqs. (2) and (3) are assumed to be true, we should expect $\chi \sim \nu \sim L^{{\alpha}_0}$, for $\phi$ constant; however, our data indicate a significantly reduced value for the exponent ${\alpha}$: the experimental values of ${\alpha}$ in the four groups of data examined in this work are of the order of $0.5\,{\alpha}_0$ to $0.7{\alpha}_0$, for G1 and G3, to $0.3\,{\alpha}_0$ to $0.5\,{\alpha}_0$, for G2 and G4. Our overall estimate is that the exponent $\alpha$ in this type of experiment can assume values in the interval $1/3$ to $2/3$, irrespective the group of data considered. We note in passing that Eqs. (2) and (3) imply that $\chi$ would be independent of $L$ if $\phi \sim L$, i.e. if the aspect ratio of the cylinder is held fixed, a result which can be experimentally tested. An important observation on Eq. (3) is that it is expected to be true only for simple boundary conditions, for instance, for a rod with fixed extremities or a rod with free extremities on a perfectly smooth non-interacting plane. It is natural to consider how the number of modes are modified by the roughness that is present in most real surfaces. For a rough cylinder of length $L$ interacting with a rough fractal plane, $\nu$ can scale as $L^{\alpha}$, where $\alpha$ is a non-trivial exponent smaller than the unit. This can occur, because in this case a fraction of the vibration modes will be forbidden or locked as a consequence of geometric engagements associated with the real topography of the interface. 

Alternately, Figs. {\ref{fig:T_N(L)_G1}} to {\ref{fig:T_N(L)_G4}} indicate that the average time between sliding events for a cylinder of length $L$, $\tau \equiv T_N(L)/N = {\chi}^{-1}$ decreases as the power law $\tau \sim L^{-{\alpha}}$, with $\alpha$ varying typically from $1/3$ to $2/3$. It is interesting to notice that this behavior is reminiscent from the observed dependence of the average interval between earthquakes ${\tau}_{\mathrm{e}}$ on a fault with length $L$: ${\tau}_{\mathrm{e}} \sim L^{-0.6}$ \cite{turcotte}. Thus the interval between earthquakes on a specified fault is longer for smaller faults, and decays as a power-law, as similarly observed in our experiments for sliding events of rough solid cylinders on rough inclines.

Now we will introduce in the discussion what seems to be the second physical aspect to be considered to explain the experimental results reported in this work: the effect of the roughness of the cylinder-chute interacting surfaces. We conjecture that the origin of the tendency of ${\chi}(L)$ for interacting rough surfaces to display power laws in $L$ in several physical situations as exemplified in Figs. {\ref{fig:T_N(L)_G1}} to {\ref{fig:T_N(L)_G4}} is connected with the frequently observed fractal {\it{nonstationary}} behavior of the roughness of surfaces \cite{liu,feder,sayles}; by this we understand that a sample of finite length $L$ taken from a real surface will never, however long, completely represents its properties. If the height $h(x)$ of a surface as a function of the position along a particular cut, $x$, is measured, the associated roughness can be defined by the width $w={\left({\left<{h(x)}^2\right> - {\left<h(x)\right>}^2}\right)}^{1/2}$, where $\left<h(x)\right> = (1/L) \int^L_0 h(x)\,dx$, and $\left<{h(x)}^2\right>=\int^L_0 h(x)^2\,dx$. A great deal of experimental data \cite{sayles,thomas,bouchaud,maloy} corroborate a robust scaling relation for $w(L)$ along an impressive interval of variability of {\it{eight}} decades in $L$, from typically $10^{-6}$m to $10^2$m:
\begin{equation}
w \sim L^{\zeta} \sim L^{3-D},
\end{equation}
where $\zeta = 3-D$ is the so called roughness exponent, and $D$ is the surface fractal dimension in physical space \cite{feder,bouchaud}. If we use the scaling hypothesis
\begin{equation}
{\chi}(L) \sim w(L),
\end{equation}
which says that the facility to generate sliding events as measured by $\chi$ is proportional to the degree of roughness of the interacting fractal nonstationary surfaces (low(high) roughness $\Rightarrow$ low(high) sliding susceptibility), the exponent $\alpha$ obtained in our experiments must be identified with the roughness exponent $\zeta$. The reader can notice that Eq. (5) does not work for Euclidean rough surfaces, as for instance one with a regular saw tooth profile. In this last case, the roughness is stationary, $w$ is independent of $L$ (i.e. $\zeta = 0$), and as $w$ grows, $\chi$ is expected to decrease. Experimental data indicates that real surfaces have $\zeta \approx 1/2$ for a multitude of natural and artificial surfaces, including concrete runways, hip joints, ship hull plates, and many types of machined metal surfaces \cite{liu,feder,sayles,thomas}, a result compatible with the exponent $\alpha$ obtained for G1 and G4 (see Figs. {\ref{fig:T_N(L)_G1}} and {\ref{fig:T_N(L)_G4}}). Furthermore, the roughness of ductile cracks of aluminum alloys that received different heat treatments is also described by Eq. (4) with a roughness exponent about $\zeta = 0.8$ \cite{bouchaud}. Another experimental study of the roughness of brittle cracks of Al-Si alloys, steel cooled by liquid nitrogen, graphite, porcelain, and other materials equally indicates the validity of Eq. (4) with $\zeta$ close to $0.8$ \cite{maloy}. From Eqs. (4) and (5), and from the numerical values of the susceptibility exponent $\alpha (\leftrightarrow \zeta)$ obtained in Figs. {\ref{fig:T_N(L)_G1}} to {\ref{fig:T_N(L)_G4}}, we conclude that the fractal dimension of the chute-cylinder interacting surfaces in our experiments are possibly random Brownian surfaces, whose fractal dimension is $D_B = 2.5$, in agreement with the data reported in Ref. \cite{sayles}. Why indeed should the sliding susceptibility be proportional to the roughness of the surface? There are in principle three arguments in favor of the conjecture given in Eq. (5):

First of all, the susceptibility is according to Eq. (1) an average on fluctuating sliding events. These fluctuations reflect obviously the fluctuating forces on the cylinders, which are in turn contact forces operating in the fractal nonstationary interface cylinder-chute. We should expect that these contact forces are affected by basic geometric aspects of the surfaces, and in particular by the surface roughness. Secondly, Eq. (5) reduces to the correct limit $\chi = 0$, for a perfect surface ($w=0$). This is expected because in this case the chute-cylinder interacting surfaces are firmly welded in a single structure due to strong atomic interactions. 
Thirdly, if we consider what is said in the previous paragraph, and apply Eqs. (4) and (5) for earthquakes, we obtain $\chi \sim L^{0.4}$ if we use the reported value $D=2.6$ for the fractal dimension of fault surfaces \cite{turcotte}. The last scaling is equivalent to say that the predicted average time between earthquakes on a fault with length $L$ would scale as ${\tau}_{\mathrm{e}} \sim L^{-0.4}$, which is close to the reported value ${\tau}_{\mathrm{e}} \sim L^{-0.6}$, obtained from seismic data \cite{turcotte}.

Moreover, if we use consistently Eqs. (2), (4) and (5), we obtain that the number of vibration modes accessible to the cylinder on the chute scales as
\begin{equation}
\nu \sim L^{3-D},
\end{equation}
a result that reduces indeed to Eq. (3) only in the Euclidean limit $D \rightarrow 2$. In principle, we expect in conformity with Eq. (6), that cylinder and chute rough surfaces with increasing values of $D$ tend to be more tightly engaged as a consequence of the increment in the effective area for microscopic interactions. This effect materializes itself in a microscopic reduction in the number of allowed vibration modes of the cylinder, i.e. $\nu$ scales with $L$ with an exponent necessarily smaller than the unit. In particular, a reduction in the number of allowed vibration modes is expected to be maximum if the interacting rough surfaces are space-filling surfaces with surface fractal dimension $D \rightarrow 3$. In this case, the surface interactions are saturated in the sense that they behave as volume-interactions that constrain the number of vibration modes to assume small values that are essentially independent of $L$ (or perharps have a logarithmic dependence on the length of the cylinder). For Brownian surfaces ($D=5/2$), we obtain the result $\chi \sim \nu \sim L^{1/2}$, which seems to describe the body of our experimental data. 

\section{Conclusions}
We have introduced a statistical response function $\chi$ that measures the degree of propensity of a rough cylinder to slide on a perturbed rough surface inclined below the angle of repose. This function was investigated from the experimental point of view and we have found that $\chi$ increases with the length of the cylinder as a nontrivial power-law in a number of situations. A possible origin for this scaling behavior is discussed in terms of a connection with the number of vibration modes of a rough cylinder on a rough chute, as well as in terms of the statistics of the roughness of the interacting surfaces. 

It is natural to ask on the influence of the several parameters appearing in the experiments on the exponent $\alpha$. Although a final answer to this subject can not be advanced at the present stage, some considerations can be outlined: (i) In the light of Eqs. (4) and (5), the reduction observed from ${\alpha}_1 = 0.55 \pm 0.05$ to ${\alpha}_2 = 0.33 \pm 0.04$ could be attributed to an effective increase in the surface fractal dimension of the cylinders after the use of the steel wool as described in Sec. 2. (ii) The increase observed from ${\alpha}_2 = 0.33 \pm 0.04$ to ${\alpha}_3 = 0.67 \pm 0.06$ could be associated to the corresponding increase in the reduced angle of inclination from ${\theta}_{\mathrm{r}} = 0.26$ to ${\theta}_{\mathrm{r}} = 0.37$. It can be noticed that the largest exponent (${\alpha}_3$) is obtained in the experiment with the largest ${\theta}_{\mathrm{r}}$. (iii) The proximity between ${\alpha}_1$ (massive cylinder) and ${\alpha}_4$ (hollow cylinder), both with almost the same ${\theta}_{\mathrm{r}}$, could suggest that the response of hollow cylinders perturbed by a light hammer is equivalent to that of massive cylinders perturbed by a heavy hammer. However, due to the very time consuming nature of the experiments described in this paper, a number of experimental aspects remain to be systematically investigated in the future. Thus, new experiments exploring the variation of the intensity of the perturbation on the chute and the variation of the materials involved, as well as the effect of the diameter of the cylinders are necessary to clarify this fluctuation phenomenon associated with the stick-slip dynamics. Furthermore, other important aspects to be answered are the type of variation of $\chi$ with the reduced angle ${({\theta}_{\mathrm{c}}-{\theta})}/{\theta_{\mathrm{c}}}$, for $L$ fixed; and the possibility of a universal value $\alpha = 1/2$ for the susceptibility exponent in any experiment involving sliding of rough solids of the type discussed in this paper.

\begin{ack}

We acknowledge H. J. Herrmann for helpful comments. We are also grateful to M. Lyra and G. L. Vasconcelos for stimulating remarks. One of us (EJRP) acknowledges a fellowship from CAPES - Bras\'{\i}lia/Brazil.

\end{ack}

\end{document}